\documentclass[conference]{IEEEtran}
\IEEEoverridecommandlockouts
\usepackage{cite}
\usepackage{amsmath,amssymb,amsfonts}
\usepackage{algorithmic}
\usepackage{graphicx}
\usepackage{textcomp}

\usepackage{booktabs} 
\usepackage{url}
\usepackage{amsmath,amssymb,amsfonts}
\usepackage{algorithmic}
\usepackage{graphicx}
\usepackage{textcomp}
\usepackage[dvipsnames]{xcolor}
\usepackage{bm}
\usepackage{stfloats}
\usepackage{multirow}
\usepackage{colortbl}
\usepackage{subfigure}

\usepackage[linesnumbered,ruled,vlined]{algorithm2e}

\SetCommentSty{mycommfont}
\SetAlCapFnt{\small}
\definecolor{codegreen}{rgb}{0,0.6,0}
\definecolor{codeblue}{rgb}{0,0,1}
\definecolor{codegray}{rgb}{0.5,0.5,0.5}
\definecolor{codepurple}{rgb}{0.58,0,0.82}
\definecolor{codered}{rgb}{1,0,0}
\definecolor{backcolor}{rgb}{0.98,0.98,0.98}
\let\oldnl\nl
\newcommand\nonl{%
    \renewcommand{\nl}{\let\nl\oldnl}}

\def\BibTeX{{\rm B\kern-.05em{\sc i\kern-.025em b}\kern-.08em
    T\kern-.1667em\lower.7ex\hbox{E}\kern-.125emX}}
\begin{document}




\title{IoTGAN: GAN Powered Camouflage Against Machine Learning Based IoT Device Identification}
\author{
    \IEEEauthorblockN{Tao Hou\IEEEauthorrefmark{2}, Tao Wang\IEEEauthorrefmark{3}, Zhuo Lu\IEEEauthorrefmark{2}, Yao Liu\IEEEauthorrefmark{2} and Yalin Sagduyu\IEEEauthorrefmark{4}}
    \IEEEauthorblockA{
      \IEEEauthorrefmark{2}University of South Florida, Tampa, FL, USA, \{taohou@, zhuolu@, yliu@cse.\}usf.edu\\
      \IEEEauthorrefmark{3}New Mexico State University, Las Cruces, NM, USA, taow@nmsu.edu\\
      \IEEEauthorrefmark{4}Intelligent Automation Inc, Rockville, MD, USA, ysagduyu@i-a-i.com\\
    }
}

\maketitle

\begin{abstract}
With the proliferation of IoT devices, researchers have developed a variety of IoT device identification methods with the assistance of machine learning. Nevertheless, the security of these identification methods mostly depends on collected training data. In this research, we propose a novel attack strategy named {\it IoTGAN} to manipulate an IoT device's traffic such that it can evade machine learning based IoT device identification. In the development of IoTGAN, we have two major technical challenges: (i) How to obtain the discriminative model in a black-box setting, and (ii) How to add perturbations to IoT traffic through the manipulative model, so as to evade the identification while not influencing the functionality of IoT devices. To address these challenges, a neural network based substitute model is used to fit the target model in black-box settings, it works as a discriminative model in IoTGAN. A manipulative model is trained to add adversarial perturbations into the IoT device's traffic to evade the substitute model. Experimental results show that IoTGAN can successfully achieve the attack goals. We also develop efficient countermeasures to protect machine learning based IoT device identification from been undermined by IoTGAN.
\end{abstract}


\section{Introduction}
The Internet of Things (IoT) refers to the network of physical devices that are embedded with sensors, chips, operating systems and other technologies, collecting and exchanging data over Internet \cite{iotwiki}. The popularization of universal computer chips and the ubiquity of wireless networks enable the revolution to transform traditional devices into smart devices as a part of the IoT. These IoT devices can be extensively deployed for different purposes including consumer (e.g., smart home, health care), commerce (e.g., transportation, manufacturing, agriculture), and military (e.g., battlefield equipment, autonomous reconnaissance). However, the heterogeneity of these devices also imposes security challenges to the management of IoT networks.

For a network containing different kinds of IoT devices, it is vital to identify the type of each IoT device before applying fine-graded security policies. Other than managing different kinds of IoT devices locally by the entity which the devices belong to, knowing the type of the device can enable a global management for security purpose from the level of the whole network, in order to permit or prohibit IoT device's specific behavior. For example, in a military base, the network should keep the geographical information confidential and forbid the surveillance camera transferring video data to the outside. Another example is that an organization may have different permissions for its personnels to access different smart devices (e.g., the maintenance staff can adjust the air conditioner unit; the security guard can view the monitor video; and any person should be able to control the smart bubble). More importantly, IoT device identification can facilitate detecting vulnerable IoT devices and preventing malicious rogue IoT devices.

Researchers have proposed various methods for IoT device identification. A simple way is using identifiers (e.g., MAC addresses, IP addresses, Bluetooth ID, Zigbee ID) to identify IoT devices. However, various identifier spoofing attacks \cite{chen2007detecting, shi2020generative} have been exploited to deceive the identification. It is necessary to develop new methods that can avoid using these identifiers for IoT device identification. Recent progresses including using statistical features which can reflect the behavior of a specific IoT device are developed for identification \cite{ammar2019network}. However, due to the limitation of mathematical model specific analysis, it is hard for these methods to keep a steady accuracy rate when applying to different real-world scenarios. Moreover, these methods usually introduce a high overhead and fail when the traffic is encrypted.

To overcome these weaknesses, researchers have introduced multiple machine learning based methods to assist IoT device identification or network traffic fingerprinting \cite{s21082660, 8116438, 8664655, 8761559, hou2020proto, hou2021combating, 8981946, kotakIoT, 9148110, leroux2018fingerprinting}. Specifically, these machine learning based methods can successfully identify IoT devices even when the identifier is spoofed, meanwhile they can achieve a high accuracy rate no matter whether the traffic is encrypted. These methods can be usually formalized as feature-based machine learning classification problems. k-Nearest Neighbors (k-NN), Support Vector Machine (SVM), Random Forest and Neural Networks are the most frequently used approaches to build the machine learning model. Through learning on traffic features on packets (e.g., packet size, header information, even encrypted content in data payload \cite{hou2019smart} ) or flows (e.g., packet interval, packet count in a time window), these machine learning models can achieve an accuracy rate as high as 99\% in IoT device identification.

Nevertheless, machine learning based IoT device identification methods are designed with the aim for high performances, but lacks any security guarantee. We discover that there exists a subtle attack surface that can undermine the machine learning based IoT device identification. Specifically, these methods rely on learning from the features which are obtained from network traffics completely or partly controlled by IoT devices. Thus, a rogue IoT device can maliciously alter its network traffic to evade the machine learning based identification. In this research, we aim to investigate current machine learning based IoT device identification methods, reveal the potential attacks, and derive corresponding countermeasures to protect the IoT device identification from been undermined.

To this end, we develop an attack strategy, named {\it IoTGAN}, which can efficiently disturb the machine learning based IoT device identification. IoTGAN is inspired by Generative Adversarial Network (GAN) \cite{goodfellow2014generative} and it allows rouge IoT devices to manipulate its traffic to camouflage themselves from been identified. We implement IoTGAN as a practical system to launch this attack. However, IoTGAN can not be simply implemented by directly using GAN. Two major technical challenges must to be addressed to achieve the malicious goal of evading IoT device identification: (i) How to obtain the discriminative model in black-box settings, and (ii) How to add perturbations to IoT traffic through the manipulative model (i.e., the generative model in GAN) to evade the identification while not influencing the functionalities of IoT devices.

In addition to the attack strategy design and analysis, we develop an effective defense approach, named {\it Device Profiling}, to protect the machine learning based IoT device identification from been undermined by IoTGAN. {\it Device Profiling} utilizes the raw wireless signals emitted from IoT devices, which exhibits inherent hardware features and cannot be manipulated by IoT devices, to mitigate the effect of IoTGAN.

We conduct experiments on real-world IoT devices with different machine learning based identification methods to evaluate the effectiveness of IoTGAN. The experimental results show that IoTGAN can evade all the identification methods with a successful rate higher than 90\%. We also conduct experiments to evaluate the defense approach and observe that the attack successful rate significantly drops to nearly zero after the deployment of Device Profiling.

The remainder of this paper is as follows. In Section~\ref{Sec:Pre}, we introduce preliminaries. In Section~\ref{Sec:iotgan}, we investigate the existing machine learning base IoT device identification methods, and state our attack strategy of IoTGAN. In Sections~\ref{Sec:disc} and \ref{Sec:mani}, we introduce the two core technical contributions in the implementation of IoTGAN. We discuss potential countermeasures to improve security in Section~\ref{Sec:count} and present the experiment results in Section~\ref{Sec:expe}. Finally, we conclude the paper in Section~\ref{Sec:con}.

\section{Preliminaries}\label{Sec:Pre}
We introduce the preliminaries in this section, including the architecture of IoT and general IoT devices.
\subsection{IoT Architecture}
Organizations like The IEEE Standards Association, International Electrotechnical Commission, and American National Standards Institute are working on developing the standards for IoT. Generally, IoT follows a multi-layer architecture.

\vspace{0.1cm}\begin{itemize}
\item {\bf Physical layer:} This layer includes the low level hardware components, such as sensors, actuators and RFIDs. IoT relies on these essential components to perform the fundamental functionalities (e.g., monitor the environment, collect information, manage operations).
\vspace{0.1cm}\item {\bf MAC/link layer:} This layer connects different devices to a network, thus to enable transmitting or exchange of the data which is collected from physical layer. The connectivity can be achieved through different kind of protocols, including WiFi, NFC, Bluetooth, ZigBee and cellular networking.
\vspace{0.1cm}\item {\bf Network layer and above:} The network layer and above connect IoT devices together via networking and provide application-level services to the end user. There are more than hundreds of applications in the IoT ecosystems, such as smart home, smart transportation, and smart city.
\end{itemize}
\vspace{0.1cm}

\subsection{IoT Devices}
An IoT network may be connected with heterogeneous devices for different applications. We introduce the mainstream IoT products in current marketplace, with the application domain they belong to and the adopted communication protocols in Table~\ref{Tab:iot}. We can find that WiFi and bluetooth are commonly used in smart home applications. While applications of smart transportation and smart city prefer cellular networks for long distance connection.

\begin{table}[!htbp]
\renewcommand\arraystretch{1.15}
\caption{Mainstream IoT products.}\label{Tab:iot}
\begin{tabular}{p{2.6cm}<{\centering}|p{1.65cm}<{\centering}|p{3.3cm}<{\centering}}
\toprule[1pt]
\rowcolor[gray]{0.9}{{\bf Name}}&{{\bf Application}}&{{\bf Communication Protocol}}\cr
\midrule
Google Nest & \multirow{4}{*}{Smart Home} & Bluetooth, WiFi\\ \cline{0-0}\cline{3-3}
Apple AirTag &  & Bluetooth, UWB, NFC\\ \cline{0-0}\cline{3-3}
Amazon Echo &  & Bluetooth, WiFi, Zigbee\\ \cline{0-0}\cline{3-3}
Samsung SmartThings &  & WiFi, Zigbee, Z-Wave\\ \cline{0-0}\cline{2-3}
Drone & \multirow{2}{*}{Smart} & WiFi, Cellular, Telemetry\\ \cline{0-0}\cline{3-3}
Smart car &  \multirow{2}{*}{Transportation} & Bluetooth, WiFi, Cellular\\ \cline{0-0}\cline{3-3}
Rail detector &  & LoRa, Cellular\\ \cline{0-0}\cline{2-3}
Smart trash& \multirow{4}{*}{Smart City}  & WiFi, Zigbee\\ \cline{0-0}\cline{3-3}
Weather station &  & WiFi, LoRa, Cellular\\ \cline{0-0}\cline{3-3}
Smart street light &  & Cellular\\ \cline{0-0}\cline{3-3}
Gunshot detector &  & WiFi, Cellular\\
\bottomrule[1pt]
\end{tabular}
\end{table}

\section{Evading ML-based IoT Device Identification} \label{Sec:iotgan}
To evade identification, it is essential for the attacker to have the knowledge of the target machine learning models. Therefore, in this section, we first investigate existing machine learning based IoT device identification methods, then we introduce the attack model to achieve the adversarial goal.

\subsection{Existing IoT Device Identification Methods}

Remote service creates a subtle attack surface to infer the identity of different IoT devices. As most IoT devices request remote service via the RESTful APIs, which adopt the uniform interface to improve the visibility of interactions, the attackers are able to learn their identities by exploring unique features extracted from uniform headers of the service request. The work in~\cite{s21082660} shows that the ML classifiers (i.e., SVM, and Logistic Regression) can reach high accuracy when the header features are considered (e.g., port numbers, domain names, and cipher suites).

Nevertheless, modeling remote service requests may not always yield sufficient information for accurate device identification, especially when the device is communicating with an anonymous service provider. More information can be gathered by learning the traffic/data flow patterns during the interaction. In particular, multiple works have been done to learn the spatial and temporal patterns of traffic/data flow to improve the identification accuracy. The work in~\cite{8116438} has collected and characterized the statistical attributes of traffic traces over 20 types of IoT devices and achieved a detection rate of 95\%. The work of \cite{8664655} models the periodic communication traffic using fingerprints extracted from frequency domain and adopts the k-NN classifier with the detection rate of 98.2\%. The authors in~\cite{8761559} automate the process of feature extraction via genetic algorithm and deploy various machine learning algorithms (i.e., DecisionTable, J48 Decision Trees, OneR, and PART) to increase the detection rate. The authors in~\cite{8981946} develop a multi-stage meta classifier that explores the flow-level attributes to further improve the classification accuracy based on the network traffic analysis. Deep learning models (e.g., CNN and RNN) have also been applied to achieve the fine-grain device identification. The work in~\cite{kotakIoT} converts the network payloads into image representation to fully capture the traffic details and achieves over 99\% overall average detection accuracy. The authors in~\cite{9148110} propose a hybrid supervised and unsupervised deep learning approach to enable the refined classification for both known and unknown device types.

{\it Important Features Used in IoT Device Identification:} Feature extraction is a key component in machine learning to have an accurate device identification. Here we summarize the most commonly used features in IoT device identification and place them into two categories: (1) Remote service features, including service request interval, service volume, service type (e.g., NTP, DNS, Storage), service domain name, and service active/sleep cycles. (2) Network packet/flow features, including local port, remote port, local address, remote address, encryption algorithm, packet size, packet interval, and communication protocol.

\subsection{Attack Model}
Machine learning based IoT device identification systems usually are hosted on the network administration side. For the attacker, the target system is a black box without disclosing the knowledge of the model internal structure and features used for identification. In this research, we aim to develop a camouflage attack that can help IoT devices to evade the machine learning based identification in a black-box setting. Though the black-box setting makes it more challenging to launch the attack, it will promote the attack to be more practical for real-world scenarios.

Specifically, we proposed a system, named IoTGAN, to achieve this attack goal. IoTGAN is inspired by GAN \cite{goodfellow2014generative}. GAN is comprised by two models: the discriminative model and the generative model. The generative model is trained to generate new samples by adding noise to the input data, while the discriminative model is used to distinguish between generated samples and real samples. Following a two-player minimax game, the generative model will be continually updated until the game achieves equilibrium.

\begin{figure}[htbp]
   \centering
   \includegraphics[width=3.4in]{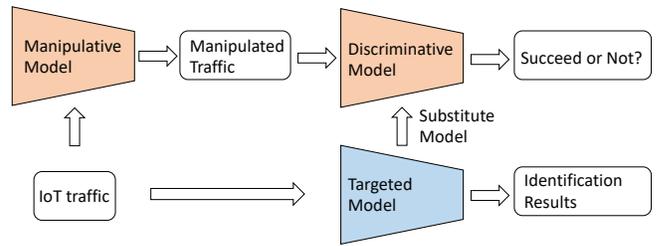}
   \caption{The architecture of IoTGAN.}
   \label{fig:iotgan}
\end{figure}

As Figure~\ref{fig:iotgan} shows, IoTGAN includes the discriminative model and the manipulative model. The discriminative model has the same functionality as in GAN to help improve the manipulative model until the manipulated IoT device traffic can evade the identification. While the manipulative model works as the generative model in GAN to manipulate the traffic of IoT devices by adding perturbations. In next two sections, we introduce our core technical contributions in the implementation of IoTGAN: (i) How to obtain the discriminative model in a black-box setting, and (ii) How to add perturbations to IoT traffic through the manipulative model to evade the identification without affecting the functionality of IoT devices.

\section{Obtaining The Discriminative Model in A Black-box Setting}\label{Sec:disc}
In this section, we introduce our design of the discriminative model in IoTGAN. The discriminative model should have the same population distribution as the target IoT device identification model, i.e., for the same input, it should have the same identification results. We aim to make it practical and can be broadly applied to a generic situation (i.e., a black-box setting in which the attacker has no knowledge of the target system). In what follows, we first present a general formula of IoT device identification. Then, we design our algorithm to obtain the discriminative model.

\begin{figure}[t]
   \centering
   \includegraphics[width=3.15in]{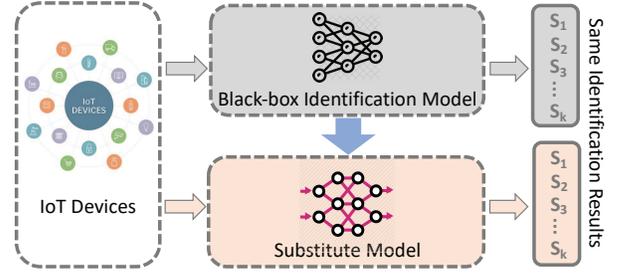}
   \caption{Obtaining a substitute model as the discriminative model in IoTGAN.}
   \label{fig:substitute}
\end{figure}

According to the literature, IoT device identification algorithms are either statistical or machine learning based methods given users' traffic features. We denote the target black-box identification algorithm as a multi-class classifier $M$. The inputs of $M$ are $K$ traffic features of a IoT device denoted as ${\mathcal H}=\{{\bf h}_1, {\bf h}_2, ... , {\bf h}_K\}$, where ${{\bf h}_i}$ is the $i^{th}$ traffic feature of the device. Accordingly, the identification results $\mathcal S$ can be written as
\begin{equation}\label{bbmodel}
\mathcal S = M(\mathcal H).
\end{equation}
The output is the identified IoT class denoted as ${\mathcal S}=\{S_i\}, i \in [1, 2, ..., N]$, where $N$ is the total classes of IoT devices that can be identified. The identifier aims to estimate the likelihood probability $P(S|H)$, which is the probability of the IoT class $S$ given the traffic input $H$. As shown in Figure~\ref{fig:substitute}, our goal is to establish a substitute model, namely, $\hat M$, adapted to the input-output relation of the target black-box model $M$, such that $\hat M$ and $M$ have the same identification results for the same group of IoT devices. This problem in fact falls into the area of model transferability \cite{papernot2016transferability} in machine learning. When labeled training datasets are collected from the same population distribution of the target black-box model, it is feasible to train a substitute model even when it has different internal structures \cite{szegedy2013intriguing}. Without loss of the generality, we adopt multi-layer full-connected neural network to learn the target black-box identifier. Sigmoid function is applied at the last layer and the class associated with the maximum probability will be selected as the output. 

\subsection{Training Data Collection}

Due to the broadcast nature of wireless communications, we assume an attacker can eavesdrop on the traffic of IoT devices and observe the identification results from the network administrator. We then extract the features from the IoT traffics and treat them as the inputs for the substitute model $\hat M$. The identification results from the network administrator will be treated as the output. As discussed, the attacker has no knowledge of traffic features used in the black-box model. To address the issue, we surveyed existing ML based IoT device identification methods and build a feature pool that contains all common traffic features used in these models. The feature pool will be used as the start point of model training. 

\subsection{Substitute Model Training}

Our model training takes two steps, 1) to obtain a substitute model that can yield the same results as the target identifier, 2) to mitigate the performance overhead by selecting a refined subset from the feature pool. 

The process to obtain the accurate substitute model can be stated as finding an $\hat M$ that minimizes the empirical loss $L$ over the training dataset. Specifically, the process can be formulated as
\begin{equation}\label{SI-goal}
\hat M = \arg\min L(\mathcal S, \hat {\mathcal S})=\arg\min\limits_{\hat M} L(\mathcal S, {\hat M(\mathcal H)}),
\end{equation}
where $\hat {\mathcal S}$ is the predicted identification result yielded by the substitute model $\hat M$ given input $\mathcal H$. The obtained substitute model will work as the discriminative model in IoTGAN.

The first step focuses on building an accurate substitute model, but it may incur considerable amount of computational overhead as hundreds of features are used in model training. To this end, IoTGAN carefully shrinks the parameter space by selecting a refined subset where only features that indeed affect accuracy will be maintained from the pool. In particular, we adopt a weight training algorithm to obtain weights for all features in the pool. Then, we create the subset by selecting top $L$ weighted features and retrain the labeled data. In order to determine the optimal size of the subset, we further define a metric Performance Gain to evaluate the prediction results of the substitute model given different values of $L$.
\begin{displaymath}
\text{Performance Gain} =  \frac{(r_c - r_p)/r_c - (c_c - c_p)/c_c}{(c_c - c_p)/c_c}.
\end{displaymath}
where $r_c$ and $r_p$ represent the accuracy rate for current and previous subset size of $L$ respectively, and $c_c$ and $c_p$ denote the computational overheads for current and previous subset size of $L$ respectively. In particular, the computational overhead is defined as the time required to identify the class of IoT device using the substitute model. A positive Performance Gain indicates that the accuracy grows faster than the computation overhead with the increasing size of the subset. In IoTGAN, $L$ is chosen empirically according to the performance gain. Specifically, we aim to keep a refined feature subset that can obtain a substitute model with high accuracy rate but at the same time impose negligible time overhead on device identification. 

\section{Manipulating the Traffic of IoT Device to Evade Identification}\label{Sec:mani}
In this section, we design the manipulative model in IoTGAN. The manipulative model is used to manipulate IoT device's traffic such that the discriminative model can not successfully identify the IoT devices. As we introduced in aforementioned content, the discriminative model in IoTGAN is a substitute model of the target identification model, therefore IoTGAN just needs to manipulate IoT device traffic to evade the obtained substitute model. In particular, the manipulate model is trained following the strategy of training the generative model in GAN. This training procedure minimizes the probability of correct identification in the discriminative model. The training process of the manipulative model in IoTGAN is shown in Algorithm~\ref{a1}.

\begin{algorithm}[h]
\small
   \SetKwInOut{Input}{Input}\SetKwInOut{Output}{Output}
    \Input{Original IoT device traffic feature vector ${\mathcal H}$}
   \Output{Trained manipulative model}
   \While(){not satisfy (2)}{Learning on original traffic ${\mathcal H}$; \\ Update the substitute model ${\hat M}$;}  \tcc{Obtain the substitute model.}
    \While(){Not converging}{Initialize the multiplier factor ${\bf r}$;\\
   Set the adversarial perturbation ${\bf s} = {\bf rh}$;\\
   \tcc{Get the perturbation.}
   Generate the manipulated traffic ${\bf h'} =G_{\theta}({\bf h}, {\bf rh})$;\\
   Label ${\bf h'}$ using the substitute model $\hat M$;\\
   Update the manipulative model's parameter set $\theta$.}
   \tcc{Train the manipulative model.}

 \caption{\small Training process of the manipulative model in IoTGAN.}\label{a1}
\end{algorithm}

The manipulative model takes IoT device traffic ${\bf h}$ and a noise ${\bf s}$ as input. The noise is a vector of traffic features. Specifically, we only consider to manipulate the features that do not affect the functionality of IoT devices. For example, the content in data payload of a packet will not be changed to avoid disturbing the communication of IoT devices. As most features used in IoT identification (e.g.,service request interval, service volume, local port, remote port) does not associate with the content of a message, it is feasible to fool the discriminative model by manipulating the traffic features. In IoTGAN, the manipulative model $G$ generates the manipulated traffic ${\bf h'} $ by
\begin{equation}
{\bf h'} =G_{\theta}({\bf h}, {\bf rh}),
\end{equation}
where ${\bf r} = \{r_1, r_2, ..., r_n\}$ is the multiplier factor for generating noise ${\bf s}$, $n$ is the number of elements in ${\bf h}$. For $i \in {1, 2,..., n}$, $r_i$ is $0$ if the corresponding feature may influence the functionality of IoT device; otherwise, $r_i$ a random number sampled from a uniform distribution of [0, 0.1]; and ${\theta}$ is the parameter set of $G$. Since we consider the discriminative model as a multi-class identifier, we cannot simply adopt existing training procedure of GAN which only works for the classifier with binary decisions. In this paper, we propose a refined training process with two operation modes 1) device misidentification, 2) identity spoofing. 

\subsection{Device Misidentification}

In device misidentification, the attacker aims to manipulate the traffic features such that the discriminative model will mislabel the given datasets. The process to achieve the purpose is to find a manipulative model $G$ that can maximize the empirical loss $L$ between the original and modified outputs over the training dataset. Specifically, the process can be formulated as
\begin{equation}\label{SI-goal}
G_{\theta} = \arg\min L(\hat {\mathcal S}, \hat {\mathcal S}_{G_{\theta}}) = \arg\max_{G_{\theta}} L({\hat M(\mathcal H)}, {\hat M( G_{\theta}({\bf h}, {\bf rh})})),
\end{equation}
where $\hat {\mathcal S}_{G_{\theta}}$ is the predicted identification results of ${\hat M}$ given the modified input $G_{\theta}({\bf h}, {\bf rh})$. After training, the attacker can hide the identify of IoT devices from been identified by the network administrator. 

\subsection{Identity Spoofing}

The misidentification can hide the identify of different IoT devices but cannot camouflage them as any specified classes for malicious purpose. For example, a surveillance camera may want to pretend as a device with no sensitive data to circumventing the rigorous export control policy. Towards this objective, identity spoofing aims to generate modified features that can fool the target model and yield the specified identification outputs designated by the attacker. 

The process of identify spoofing can be formulated as finding a manipulative model $G$ to minimize the empirical loss $L$ between the specified class and outputs given the modified training dataset. Specifically, it is described as
\begin{equation}\label{SI-goal}
G_{\theta} = \arg\min_{G_{\theta}} L(S_{spoof}, {\hat M( G_{\theta}({\bf h}, {\bf rh})})),
\end{equation}
where $s_{spoof}$ is the class specified by the attacker. After training, the attacker is able to deceive the discriminative model and pretend as any type of IoT device. 

\section{Countermeasures}\label{Sec:count}

As discussed in Section~\ref{Sec:iotgan}, traffic based features can be subtly manipulated by IoT devices to evade machine learning based IoT device identification. To defend against IoTGAN, we aim to identify different IoT devices using features that cannot be easily manipulated. The proposed classifier is complementary to existing IoT identification methods and can be easily integrated with them to improve identification accuracy.

We observe that raw wireless signals emitted from different IoT devices can exhibit inherent hardware features that cannot be forged by common users. In particular, manufacturing imperfection existing in IoT devices may impose a substantial change on the transmit signal waveforms, yielding unique features for device identification. We also note that wireless signals experience distinct channel distortions when they travel through different propagation paths \cite{sagduyu2019adversarial, wang2015location, wang2016signal}. We aim to take advantage of these nonlinear characteristics of radio channels to fingerprint different devices at various locations.

Specifically, we propose a method named {\it Device Profiling} to distinguish different IoT devices. The method includes two components: 1) feature profiling, which statistically describes the nonlinear characteristics of the transmit signals from IoT devices; 2) device fingerprinting, which builds a neural network based multi-stage classifier to learn the feature patterns for accurate and efficient device identification, even when the traffic based features are manipulated.

In feature profiling, we profile different IoT devices using features extracted from radio frequency signals in four perspectives: amplitude attenuation, phase shift, frequency offset, and arrival angle. In particular, amplitude attenuation and phase shift can be extracted from channel estimation at the receiver. Both features indicate the channel distortions caused by the internal hardware imperfection and the distinct prorogation paths. Frequency offset can be estimated by the maximum likelihood algorithm to derive the frequency deviation caused by transmitters's internal imperfections. The arrival angle can be measured via the multiple-input and multiple-output (MIMO) technique and exhibits the location-specific information of propagation channels.

In device fingerprinting, we build a multi-stage classifier which combines CNN and multi-class decision tree for accurate and efficient device identification. The input of the classifier is the profiled features and the channel state information estimated at the receiver, while the output is the mapping results of different IoT devices (i.e., a possibility associated with each specific device is generated to indicate how likely the input belongs to the device). In particular, the CNN is used to learn the subtle differences between different IoT devices that cannot be fully captured by the features extracted in feature profiling. The multi-class decision tree breaks down the whole dataset into different levels and make decisions step by step, in that way we only travel through the branches with high possibility, reducing the unnecessary searching space.

Unlike traditional link signature based identification~\cite{6195669, 5601729}, where a location-specific metric is extracted from radio channels to localize devices at different positions, the proposed approach considers both internal hardware imperfection and external location-specific features to distinguish devices with different identities (e.g., types, locations). In addition, the deep learning based classifier is expected to better fuse different features and achieve a higher accuracy.

Our experiment shows that the propose Device Profiling can achieve fast and accurate IoT device identification even in the presence of the manipulated traffic flows.

\begin{table*}[htbp]
\centering
\renewcommand\arraystretch{1.15}
\caption{The identification rate for different target models and trained discriminative models.}\label{tr1}
\begin{tabular}{p{2.6cm}<{\centering}|p{2.3cm}<{\centering}|p{2cm}<{\centering}|p{2.3cm}<{\centering}|p{2cm}<{\centering}}
\toprule[1pt]
\multirow{2}*{\bf Model}& \multicolumn{2}{c|} {\bf Target Model} & \multicolumn{2}{c} {\bf Discriminative Model}\\ \cline{2-5}
&{\bf Training Dataset}&{\bf Test Dataset}&{\bf Training Dataset}&{\bf Test Dataset}\cr
\midrule
 Random Forest &  97.62\% &97.32\% &96.36\% &97.10\% \\ \cline{1-5}
 Decision Tree &  92.20\% &93.25\% &  92.12\% &90.89\% \\ \cline{1-5}
 SVM &  97.89\%  &96.58\% &  96.50\% &95.52\% \\ \cline{1-5}
 k-NN &  93.11\%  &92.33\% &  90.12\% &91.56\% \\ \cline{1-5}
 Neural Networks &  98.56\% &98.86\% &  98.00\% &97.92\% \\
\bottomrule[1pt]
\end{tabular}
\end{table*}

\section{Experimental Evaluations}\label{Sec:expe}
In this section, we evaluate the attack performance of IoTGAN and the defense effectiveness of Device Profiling.

\subsection{Experimental Setup}

The dataset used in evaluation is obtained from UNSW IoT Trace Data \cite{sivanathan2018classifying}. The dataset is collected in real-world environment deployed with 28 different IoT devices. These IoT devices include smart cameras (e.g., Samsung SamrtCam, Belkin camera, TP-Link Cloud camera, and Ring door bell), smart hubs (e.g., Amazon Echo, Samsung Smart Things), smart switches (e.g., Belkin motion detector, and TP-Link smart plug) and healthcare devices (e.g., Withings Smart scale, Blipcare blood pressure meter). In the evaluation, we split the IoT trace dataset into two parts. The first part contains 80\% of the data as the training dataset, while the remaining 20\% is used as the test dataset. In order to validate the performance of IoTGAN, we adopt several typical machine learning based IoT device identification models as the target black-box models, including Random Forest \cite{singh2019optimization}, Decision Trees\cite{s21082660}, SVM \cite{s21082660}, k-NN \cite{8664655}, and Neural Networks \cite{9148110}.

\subsection{The Evaluation of IoTGAN}
We following the references to implement the five typical machine learning based IoT device identification models. Then we use the UNSW IoT Trace Data to train and test these models. We use a metric named identification rate to evaluate the performance of IoTGAN, it is defined as
\begin{align}
\text{Identification\ Rate} = \frac{\text{Correct Identification Count}}{\text{Total Identification Count}}.\nonumber
\end{align}

The evaluation results for the five target machine learning based IoT device identification models without the attack of IoTGAN is shown in the left two columns of Table~\ref{tr1}. We can observe that all these five typical identification models can achieve a high identification rate larger than 92\% for IoT device identification. For the Neural Networks based model, the identification rate can even achieve 98.86\%.

The evaluation results of Table~\ref{tr1} indicate that the five selected typical identification models are able to successfully identify IoT devices when there is no malicious interference. Therefore, they can work as good baselines to evaluate the effectiveness of IoTGAN. If their identification rates are significantly dropped under the attack of IoTGAN, our proposed attack strategy will be proved effective in camouflaging IoT devices.

\begin{figure}[htbp]
   \centering
   \includegraphics[width=2.85in]{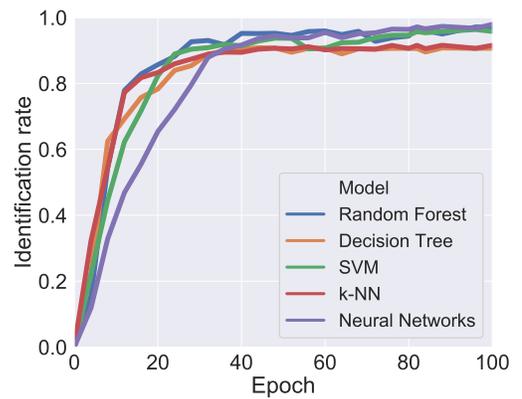}
   \caption{The identification rates for the discriminative model.}
   \label{dt}
\end{figure}

\subsubsection{The Evaluation of Discriminative Model}
We train discriminative models and manipulative models for all five target models with the UNSW IoT Trace Data. The same training and test datasets are used to obtain the discriminative models. Since discriminative models are the substitutes of the target identifiers, they should have  the similar identification rate. The right two columns of Table~\ref{tr1} demonstrate the performance of the discriminative models. As shown, our substitute models can achieve similar identification rate as the target ones. The results indicate that our proposed training process can effectively learn the target identifiers and generate the same identification results. 

Figure~\ref{dt} shows the identification rates of the discriminative models as the number of epochs increases. As shown, the discriminative models for all target identifiers can reach the stability after 40 epochs and approach an identification rate higher than 90\%. In particular, the discriminative model for random forest identifier can achieve a high identification rate of 92.10\% after 40 epochs. 

\subsubsection{The Evaluation of Manipulative Model}
We then use IoTGAN to attack these target models and calculate the new identification rate for both the training dataset and test dataset. We first mount the attack of the device misidentification to reduce the identification rate of the target model. The evaluation results are shown in Table~\ref{r2}. We can observe that IoTGAN is able to decrease the identification rates to almost zero for all five target models. This indicates that IoTGAN can successfully manipulate the IoT traffic to evade machine learning based IoT device identification.

\begin{table}[!htbp]
\centering
\renewcommand\arraystretch{1.2}
\caption{The identification rate for different machine learning based IoT devices identification models under the attack of IoTGAN.}\label{r2}
\begin{tabular}{p{2.6cm}<{\centering}|p{2.3cm}<{\centering}|p{2cm}<{\centering}}
\toprule[1pt]
{\bf Model}&{\bf Training Dataset}&{\bf Test Dataset}\cr
\midrule
 Random Forest & 0.15\% & 0.13\%\\ \cline{1-3}
 Decision Tree & 0.05\% & 0.06\% \\ \cline{1-3}
 SVM &   0.11\% & 0.08\% \\ \cline{1-3}
 k-NN &   0.21\% & 0.23\% \\ \cline{1-3}
 Neural Networks &  0.08\% &0.06\% \\
\bottomrule[1pt]
\end{tabular}
\end{table}

Next, we launch the attack of identity spoofing to camouflage IoT devices with specified types. In UNSW IoT Trace Data, we have four main types of IoT devices (i.e., smart cameras, smart hubs, smart switches and healthcare devices.). We conduct an experiment of the identify spoofing among all these types of IoT devices and evaluate their performance. In particular, we define a metric named Spoofing Rate that indicates the successful rate of the spoofing attack. It is described as the ratio between the count of successful spoofing identifications and the total count of identifications, 
\begin{align}
\text{Spoofing\ Rate} = \frac{\text{Successful Spoofing Identification Count}}{\text{Total Identification Count}}.\nonumber
\end{align}

Table~\ref{tr3} demonstrates the performance of the identify spoofing. As shown, we can achieve more than 90\% successful rate when spoofing the IoT identify between {\bf smart camera $\Leftrightarrow$ smart hub}; {\bf smart camera $\Leftrightarrow$ healthcare device}; {\bf smart hub $\Leftrightarrow$ healthcare device}; {\bf smart switch $\Leftrightarrow$ healthcare device}. Such results indicate that the attack can effectively hide the identity of IoT devices and designate them new specified ones. However, we also find two exceptions (i.e., {\bf smart camera $\Leftrightarrow$ switch}, {\bf smart hub $\Rightarrow$ switch}) that can only achieve the spoofing rate around 70\%. This may because the traffic pattern of the smart camera is quite different from the smart switch that can hardly be spoofed. Specifically, the smart camera usually enables a real-time video transmission that demands a high bandwidth and privileged wireless channel. On the other hand, the data from the smart switch is relatively static and sparse that only occupies a very limited bandwidth. Due to the nature difference of the traffic pattern between the camera and switch, it's hardly to hide some trivial features without affecting its functionalities. 

\begin{table*}[htbp]
\centering
\renewcommand\arraystretch{1.25}
\caption{The spoofing rate for different types of identity spoofing attack ({\bf camera $\Leftrightarrow$ hub} indicates the identify spoofing between the smart cameras and smart hubs. {\bf camera $\Rightarrow$ hub} indicates the attacker aims to camouflage the identify of smart camera to the smart hub.)}\label{tr3}
\begin{tabular}{p{2.1cm}<{\centering}|p{0.7cm}<{\centering}|p{0.7cm}<{\centering}|p{0.8cm}<{\centering}|p{0.8cm}<{\centering}|p{0.8cm}<{\centering}|p{0.8cm}<{\centering}|p{0.7cm}<{\centering}|p{0.7cm}<{\centering}|p{0.7cm}<{\centering}|p{0.7cm}<{\centering}|p{0.7cm}<{\centering}|p{0.7cm}<{\centering}}
\toprule[1pt]
{\bf Identity Spoofing} & \multicolumn{2}{c|} {\bf camera $\Leftrightarrow$ hub} & \multicolumn{2}{c|} {\bf camera $\Leftrightarrow$ health} & \multicolumn{2}{c|} {\bf camera $\Leftrightarrow$ switch} & \multicolumn{2}{c|} {\bf hub $\Leftrightarrow$ health} & \multicolumn{2}{c|} {\bf hub $\Leftrightarrow$ switch} & \multicolumn{2}{c} {\bf switch $\Leftrightarrow$ health}\\ \cline{1-13}
{\bf Model} & $\Leftarrow$ & $\Rightarrow$ & $\Leftarrow$ & $\Rightarrow$ & $\Leftarrow$ & $\Rightarrow$ & $\Leftarrow$ & $\Rightarrow$ & $\Leftarrow$ & $\Rightarrow$ & $\Leftarrow$ & $\Rightarrow$ \cr
\midrule
 Random Forest &  92.11\% &91.59\% &85.12\% &89.13\% & 77.12\% &68.31\% &91.13\% &90.12\% &  91.87\% &75.63\% &90.34\% &91.31\% \\ \cline{1-13}
 Decision Tree &  93.19\% &92.21\% &  90.23\% &89.35\% & 70.79\% &69.21\% &89.23\% &92.49\% &  92.85\% &71.25\% &91.13\% &92.95\% \\ \cline{1-13}
 SVM &  89.17\%  &91.75\% &  89.72\% &91.89\% & 72.65\% &70.12\% &92.54\% &92.13\% &  90.15\% &69.92\% &93.56\% &92.79\% \\ \cline{1-13}
 k-NN &  91.57\%  &91.12\% &  88.21\% &90.92\% & 68.15\% &69.78\% &91.45\% &90.68\% &  89.74\% &70.54\% &91.51\% &90.36\% \\ \cline{1-13}
 Neural Networks &  93.67\% &93.98\% &  89.10\% &87.88\% & 73.64\% &71.23\% &91.65\% &93.68\% & 91.20\% &71.99\% &92.55\% &93.79\%\\
\bottomrule[1pt]
\end{tabular}
\end{table*}

\subsection{The Evaluation of Device Profiling}

\begin{figure}[htbp]
   \centering
   \includegraphics[width=2.72in]{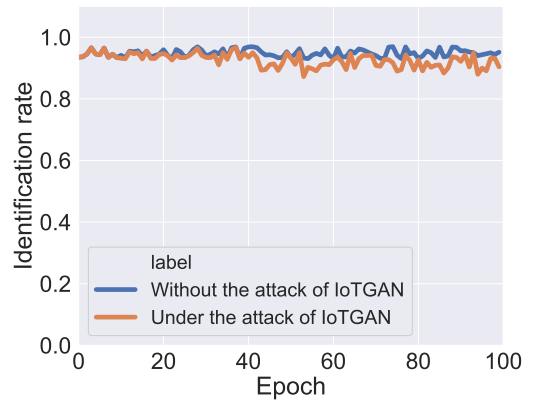}
   \caption{The identification rates for Device Profiling in the scenario: (1) without the attack of IoTGAN, and (2) under the attack of IoTGAN.}
   \label{fr1}
\end{figure}

We implement the Device Profiling following the design in Section~\ref{Sec:count}. It mainly includes two components: the feature profiling, which uses the nonlinear characteristics of the transmit RF signals for IoT device profiling; and the device fingerprinting, which build a neural network based multi-stage classifier to learn the feature patterns for accurate and efficient device identification even when the traffic based features are manipulated. The system is built with Intel WiFi Wireless Link 5300 802.11n MIMO radio tool \cite{halperin2011tool} to collect data. Device Profiling is deployed in a real-world environment with different IoT devices (e.g., Eufy security camera, Eufy smart lock, and Ecobee thermostat) for evaluation. We then use IoTGAN to attack this Device Profiling system. The evaluation results is shown in Figure~\ref{fr1}

From Figure~\ref{fr1}, we can see when there is no attack of IoTGAN, Device Profiling can achieve an identification rate around 98\%. When IoTGAN is deployed to launch attacks, there is nearly no effect on the identification rate of Device Profiling when the epoch of training the manipulative model is less than 30. This indicates that the generative model needs approximate 30 epochs of training to approach the stable state. After that, the Device Profiling can still maintain an identification rate around 95\%. Therefore, we can conclude that Device Profiling can successfully defend against the attacks launched by IoTGAN.

\section{Conclusions}\label{Sec:con}
In this paper, we investigate the security of machine learning based IoT device identification methods. We propose a novel attack strategy named IoTGAN to manipulate the IoT devices' traffic such that it can evade machine learning based IoT device identification. In IoTGAN, a substitute model is designed to fit the target identification model in a black-box setting, and a manipulative model is trained to add adversarial perturbation into IoT devices' traffic to evade the substitute model. Experimental results show that IoTGAN can effectively disrupt device identification. We also develop countermeasures complementary to existing methods to further protect machine learning based IoT device identification.

\section*{Acknowledgment} 
The work at USF was supported in part by NSF CNS-1553304 and ECCS-2029875.

\bibliographystyle{IEEEtran}
\bibliography{milcom}

\end{document}